\documentstyle[11pt,newpasp,epsfig,twoside]{article}
\def\edcomment#1{\iffalse\marginpar{\raggedright\sl#1\/}\else\relax\fi}
\reversemarginpar

\begin{document}
\def\blankline  {\vskip10truept}
\def\sal {\vskip 12 truept}
\def\lax    {\ifmmode{_<\atop^{\sim}}\else{${_<\atop^{\sim}}$}\fi}
\def\gax    {\ifmmode{_>\atop^{\sim}}\else{${_>\atop^{\sim}}$}\fi}

\title{Orbital Motion of Resonant Clumps in Dusty Circumstellar
Disks as a Signature of an Embedded Planet}
\author{Nick N. Gorkavyi}
\affil{NRC/NAS, Code 685, NASA/GSFC, Greenbelt, MD 20771}
\author{Leonid M. Ozernoy}
\affil{5C3, Computational Sciences  Institute and Department of Physics
\& Astronomy, George Mason U., Fairfax, VA 22030-4444;\\
{\rm also} Code 685, NASA/GSFC, Greenbelt, MD 20771}
\author{John C. Mather}
\affil{Code 685, NASA/GSFC, Greenbelt, MD 20771}
\author{Sara Heap}
\affil{Code 681, NASA/GSFC, Greenbelt, MD 20771}

\begin{abstract}
     We have applied a powerful numerical approach to compute,
with a high resolution, the structure of dusty circumstellar
disks with  embedded planets. We emphasize
some testable implications of our simulations which would
verify the presence of a planet via thermal emission of one or
more dusty clumps which are in  mean motion resonances
with the planet. In particular, our simulations indicate that
Vega may have a massive planet of $m\sim 2~m_J$ ($m_J$ being
Jupiter's mass) at a distance of
50--60~AU, and Epsilon Eri may have a less massive planet
of $m\sim 0.2~m_J$ at a similar
distance of 55--65 AU. This conclusion is testable: Each
resonant feature is stationary in the reference frame
co-rotating with the planet, but it is not so for the observer
at Earth. Therefore, if our interpretation of asymmetric clumps
 in circumstellar disks  as 
dynamical resonant structures is correct, the above
pattern  revolves around the star with an angular velocity of
$(1.2-1.6)^\circ$/yr (Vega) and $(0.6-0.8)^\circ$/yr
($\epsilon$~Eri) -- a prediction that can be
tested on a timescale of several years.
\end{abstract}

\section {INTRODUCTION}

Dusty disks around stars are a very common cosmic phenomenon.
These disks, with scales of tens to hundreds (up to a thousand) of
astronomical units, demonstrate a great variety in structure;
sometimes they have a central `hole' void of  gas and dust,
and often are highly asymmetric. The dusty circumstellar disks
are thought to accompany planetary systems and,
at the same time, to hide them from the observer. Knowledge
of dust characteristics is of prime importance for future
NASA space missions, such as the NGST (Next Generation Space
Telescope ) and the Terrestrial Planet Finder. It is generally
expected that a significant limitation to unambiguous planet
detection and study will be contaminating thermal emission
from dust in the target systems. We propose to turn this hazard into
an advantage for detecting and characterizing planetary systems.
Both the stellar radiation drag and stellar wind drag as well as
residual gas in the disk tend to induce dust inflow toward the star.
As dust particles pass by the planets in their infall, they interact
with them by accumulating in planetary resonances, which are observable
as clumps or belts in the dust disk.

\section {NUMERICAL SIMULATIONS}

We focus on the conditions favorable to formation of resonant
rings near the orbit of just one planet orbiting the parent star.
Recently, we have elaborated a novel, very efficient approach to numerical
modeling of distributions of test particles in an external gravitational
field (Ozernoy, Gorkavyi, \& Taidakova 2000; Gorkavyi et al. 1999;
Taidakova \& Gorkavyi 1999). Our approach removes the major obstacle 
to reliable numerical simulations -- the particle-number limitation.
 In brief, our approach (which has a number of
common elements with the `particle-in-cell' computational method) is as
follows: Let us consider, for simplicity, a stationary particle
distribution in the frame co-rotating with the planet. The locus of the
given dust particle's  positions (taken, say, as $10^4$
positions every revolution about the star) are recorded and considered as the
positions of {\it many other particles} produced by the same source of
dust but {\it at a different  time}. After this particle `dies' (as a result
of collisions, infall, or ejection from the system by a planet-perturber),
its recorded positions
sampled over its lifetime form a stationary distribution as if it were
produced by {\bf many} particles. Typically, each run includes $10^3$
revolutions, i.e.  $\sim 10^7$ positions of a dust particle, which is
equivalent, for a stationary distribution, to $10^7$ particles.
Allowing for 1000 sources of dust, we deal, after 1000 runs, with
$\sim 10^{10}$ particle positions as if they were real particles.
In the present paper, we immediately sort this information
into $\sim 10^7$ spatial cells (each cell containing $10^3-10^4$ particles),
thereby forming a 3D grid that models the dust cloud around the star. We have
computed about 300 model disks exploring the effects of various resonances
given the location of dust sources and adopting as parameters the mass of
the planet, initial orbital characteristics of dust particles, and
particle radius $r$ (in $\mu$m) coupled with stellar luminosity $L_*$
 and mass $M_*$ (in solar units), as described by
the parameter $\beta\approx 0.3~L_*/(M_*r)$.

The modelling has shown that a planet, via resonances and
gravitational scattering, produces an asymmetric resonant dust
belt with one or more clumps plus one or more
cavities as well as a central region devoid of dust.
These features can serve as indicators of a planet embedded
in the circumstellar dust disk and, moreover, can be used to
determine the mass of the planet and even some of its orbital
parameters. Our simulations in more detail, as well as related work,
are described by Ozernoy et al. (2000).

\begin{figure}[!htbp] 
\centerline{\epsfig{file=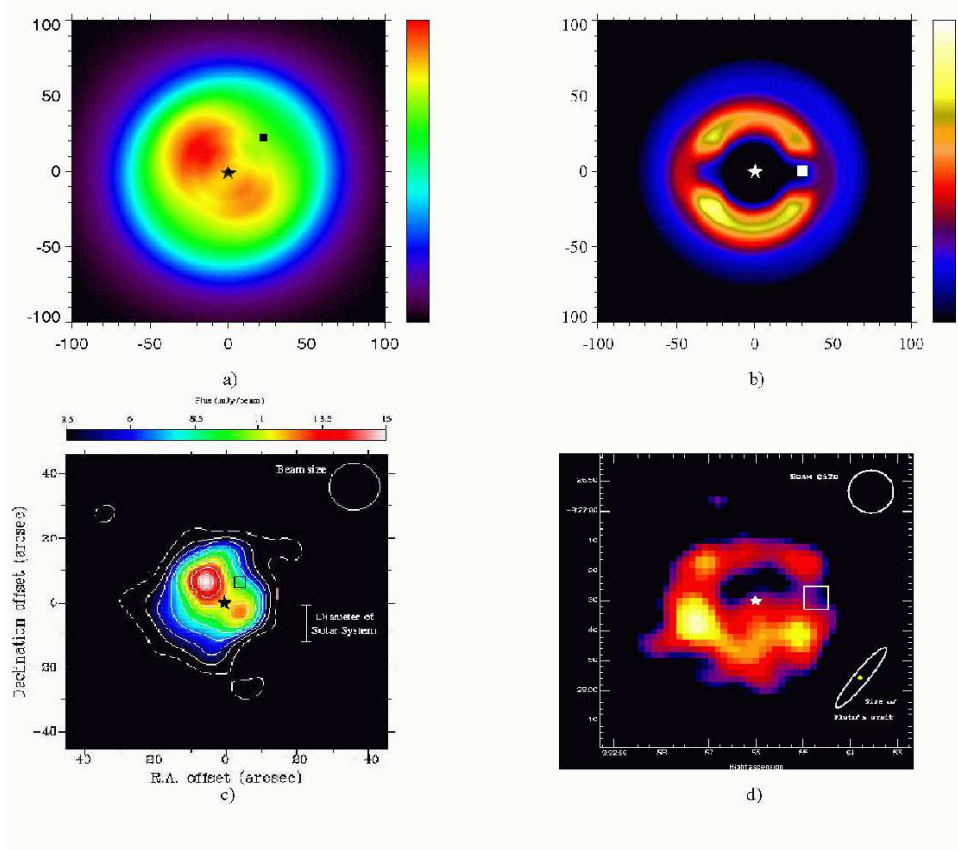,width=5.0in,height=4.46in}}
\caption{
 Thermal emission ($\lambda = 850~\mu$m) from simulated
circumstellar disks with the resonant structures vs. observations. The 
star's location is (0,0). 
The unit of length $={1\over 30}$ of the planet's radius.
The planet (shown in {\bf a, b} as a  square)  revolves, like the disk,
counterclockwise.
In  {\bf c, d}, a square indicates the expected planet's location.
To compare our simulations with observations, averaging with the
beam size (50 units of length in  {\bf a} and 20 units of length in  
{\bf b} is applied to the simulations.
%
\noindent
-- {\bf a)}
The simulation reminiscent of  the disk of Vega:
$M_\star=2.5~{\rm M}_\odot$;  $m_{pl}=2~m_J$; $\beta=0.3$; the lifetime
of dust particles $\tau=
2\times 10^4$ planet's revolutions; the number of dust sources (comets)
is 486, and the number of particle positions is $3\times 10^{10}$. The 
adopted size of the central cavity is 10 units of length. For a version 
of this simulation with the beam size of 40 and the central cavity size
of 12 units of length, see Ozernoy et al. (2000).
 -- {\bf b)}~The simulation reminiscent of the disk of $\epsilon$ Eridani:
$M_\star=0.8~{\rm M}_\odot$; $~m_{pl}= 0.2~m_J$; the resonances 2:1 and
3:2 (populated in ca. equal proportions); $\beta=0.002; \tau= 10^2$ planet's 
revolutions; the number of dust sources (kuiperoids)
is 2910 (of them, 1250 sources form the upper branch and 1660 sources
form the lower branch of the pattern), and the number of particle positions 
is $2\times 10^{9}$.
-- {\bf c)}~Circumstellar disk  of Vega (Holland et al. 1998).
-- {\bf d)}~Circumstellar disk  of $\epsilon$ Eridani (Greaves et al. 1998).
}
\label{fig1}
\end{figure}

\section {COMPARISON WITH OBSERVATIONS}

The results of this study reveal a remarkable similarity
with various types of highly asymmetric circumstellar disks observed
at submillimeter wavelengths around $\epsilon$ Eri and Vega.
Fig.~1a,b show the thermal emission from simulated disks (seen face-on)
with an embedded planet, which are to be compared with the available
observational data on Vega and  $\epsilon$ Eri shown in Fig.~1c,d (the
parent stars are seen almost pole-on). The sources of dust are assumed 
to be alike (gravitationally scattered) comets in the Vega's disk and
alike (resonant) kuiperoids in the $\epsilon$ Eri disk.
A planet as massive as 2 Jupiter masses ($m_J$)
(or somewhat less if the role of resonant comets is appreciable)
produces, from the cometary dust, two clumps (Fig.~1a), which is reminiscent of
the observed  asymmetric clumps near Vega shown in Fig.~1c.
A smaller mass $\sim 0.2~m_J$  induces, from the kuiperoidal dust,
 two asymmetric arcs, with
four clumps at its edges (Fig.~1b). A similar asymmetric structure
has been revealed in sub-mm imagery  of the $\epsilon$ Eri disk
shown in Fig.~1d.

Given the particular resonant pattern  as
the site for the captured dust, the appearance of the resonant structure 
in a dusty disk depends on the beam size and the presence/absence of a 
central cavity in the dust distribution. The latter can be produced as a 
result of gravitational scattering of the dust by a massive planet
in the inner part of the planetary system, although 
the emission of dust from inner comets  may mask its presence. 

The above modeling indicating that Vega may have a planet of mass
$\sim~2~m_J$ at a distance of $50-60$ AU, and $\epsilon$ Eri may
have a less massive planet of $m\approx 0.2~m_J$ at a similar distance
of $55-65$ AU, is testable: Each resonant feature is stationary in the
reference frame co-rotating with the planet, but it is not so for the
observer at Earth. Therefore, if the proposed interpretation of asymmetric
clumps as dynamical resonant structures is correct, the above
asymmetric feature revolves around the star
with an angular velocity of $(1.2-1.6)^\circ$/yr (Vega) and
$(0.6-0.8)^\circ$/yr ($\epsilon$ Eri) -- a prediction
that can be tested within several years.
By the middle of 2000, one could expect the resonant features to be shifted,
compared to their original positions in 1997-1998, by $4-5^\circ$ in the
Vega's disk and by $\sim 2^\circ$ in the $\epsilon$ Eri disk.
In practical terms, an
azimuthal brightness distribution at $R=50-60$ AU in the disk
is to be measured and cross-correlated on a timescale of a few years.
This could make it possible to reveal the revolution of the resonant pattern,
to determine its direction, and evaluate the rate of motion.

If confirmed, the proposed
interpretation  of the structure in Vega- and $\epsilon$ Eri- like
circumstellar disks seen face-on would make it possible not just
to reveal the embedded planet and determine its semimajor axis, but
also to constrain its other basic parameters, such as the planet's mass
and even to pinpoint the position of the planet.
Our work suggests important observational tasks as follows:

{\it A. to determine the direction of orbital revolution for dusty clumps;}

{\it B. to estimate the angular velocity of those clumps;}

{\it C. to reveal a faint peripheral structure external to the clumps.}

As our modeling indicates, an answer to task A would enable
as to eliminate an uncertainty in the position of the predicted
planet; its mass could be estimated more accurately as well.
Task B would be important to determine the planet's
orbital radius. Finally, task C would enable us to determine
which particular resonances are the strongest and to pinpoint
the position of the cavity in which the planet is located.

So far the most successful method of detecting extrasolar planets,
such as the precision Doppler measurements, is biased
toward comparatively small distances from the star. Our novel method
offers a complementary approach for revealing invisible {\it outer}
planets in circumstellar disks and determination of planetary parameters
using the visible morphology of the {\it outer} part of the disk.

\blankline
{\it Acknowledgements}. This work has been supported by NASA Grant NAG5-7065
to George Mason University. N.G. has been supported through NAS/NRC
Associateship Research program. We thank to W.S.~Holland and J.S.~Greaves
for permission to reproduce their SCUBA images.


\def\ref#1  {\noindent \hangindent=24.0pt \hangafter=1 {#1} \par}

\end{document}